\documentclass[copyright,creativecommons]{eptcs}
\providecommand{\event}{WWV 2011} 
\usepackage{breakurl}             

\usepackage{moreverb} 
\usepackage{pdfpages}
\usepackage{amsmath}

\usepackage{graphicx}
\usepackage{amsmath}
\usepackage{fancybox}
\usepackage{tikz}
\usepackage{amsfonts}
\usepackage{amssymb}
\usepackage{lscape}
\usepackage[utf8]{inputenc}

\usepackage{listings}
\usepackage{multirow}
\usepackage{multicol}
\usepackage{graphicx} 

\usepackage{url}
\definecolor{Blue}{rgb}{0.1,0.1,0.9}
\definecolor{Red}{rgb}{0.9,0.1,0.1}

\renewcommand{\cal}[1]{\mathit #1}

\newcommand{\overarrow}[1]{\stackrel{#1}{\rightarrow}}

\makeatletter
\DeclareRobustCommand\varname[1]{%
\ifmmode
  \begingroup 
  \let\math@bgroup\@empty \let\math@egroup\@empty \mathit\@empty #1\endgroup
\else
  \textit{#1}
\fi}
\makeatother

\newcommand{\raw}{\rightarrow}

\def\defemb#1#2{\expandafter\def\csname #1\endcsname
                              {\relax\ifmmode #2\else\hbox{$#2$}\fi}}

\def\ll{[\![}
\def\rr{]\!]}
\def\Den#1{\relax\ifmmode \ll #1\rr \else\hbox{$\ll #1\rr$}\fi}

\let\|=\mid

\def\#{\hat{~}}

\defemb{implies}{\quad\rightarrow\quad}
\defemb{Implies}{\quad\Rightarrow\quad}

\defemb{cA}{{\cal A}}
\defemb{cB}{{\cal B}}
\defemb{cC}{{\cal C}}
\defemb{cD}{{\cal D}}
\defemb{cE}{{\cal E}}
\defemb{cF}{{\cal F}}
\defemb{cG}{{\cal G}}
\defemb{cH}{{\cal H}}
\defemb{cI}{{\cal J}}
\defemb{cJ}{{\cal J}}
\defemb{cL}{{\cal L}}
\defemb{cP}{{\cal P}}
\defemb{cR}{{\cal R}}
\defemb{cS}{{\cal S}}
\defemb{cT}{{\cal T}}
\defemb{cU}{{\cal U}}
\defemb{cV}{{\cal V}}
\defemb{cZ}{{\cal Z}}

\DeclareSymbolFont{boldsymbols}{OMS}{cmsy}{b}{n}
\DeclareSymbolFontAlphabet{\mathbfcal}{boldsymbols}

\long\def\comment#1{}

\newcommand{\startprog}{\begin{prog}}
\newcommand{\stopprog}{\end{prog}\noindent}

\newcommand{\scripting}[2]{
{\small \sf
\begin{center} 
#1   \scriptsize
 \begin{tabular}{|l|}
  \hline			
  #2 \\
  \hline  
\end{tabular}
\end{center}  } }

\usepackage{listings}
\title{Debugging of Web Applications with {\sc Web-TLR}\thanks{
This work has been partially supported by the EU (FEDER) and the Spanish MEC TIN2010-21062-C02-02 project, 
by Generalitat Valenciana, ref. PROMETEO2011/052,
and by the Italian MUR under grant RBIN04M8S8, FIRB project, Internationalization 2004. Daniel Romero is also supported by FPI–MEC grant BES–2008–004860.
}}
\author{Mar\'ia Alpuente \qquad Javier Espert \qquad Francisco Frechina \qquad Daniel Romero
\institute{
DSIC-ELP, Universidad Polit\'ecnica de Valencia\\ 
Camino de Vera s/n, Apdo 22012, 46071\\
Valencia, Spain}
\email{\{alpuente,jespert,ffrechina,dromero\}@dsic.upv.es}
\and
Demis Ballis
\institute{Dipartimento di Matematica e Informatica \\
Via delle Scienze 206, 33100 \\ Udine, Italy}
\email{demis.ballis@uniud.it}
}
\def\titlerunning{Debugging of Web Applications with {\sc Web-TLR}}
\def\authorrunning{M. Alpuente, D. Ballis, J. Espert, F. Frechina, \& D. Romero}

\begin{document}
\maketitle

\begin{abstract}
\textsc{Web-TLR} is a  Web
verification engine that is based on the 
well-established  \emph{Rewriting Logic--Maude/LTLR\/}  tandem for   
  Web  
system specification and 
model-checking. 
In \textsc{Web-TLR}, Web applications are expressed as rewrite theories that can be formally verified by using the Maude built-in LTLR model-checker.
Whenever a property is refuted, a counterexample trace  is delivered   that reveals an undesired, erroneous navigation sequence.
Unfortunately, the analysis (or even the simple inspection) of such counterexamples may be unfeasible because of the size and complexity of the traces under examination.
In this paper, we endow \textsc{Web-TLR} with a new Web debugging facility 
that supports the efficient manipulation of counterexample traces.
This facility is based on a backward trace-slicing technique for rewriting logic theories that allows 
the pieces of information that we are interested  to be traced back through inverse rewrite sequences.
The slicing process drastically simplifies the computation trace by dropping useless data that do not influence the final result.
By using this facility,
the Web engineer can focus on the relevant fragments of the failing   application, which greatly reduces the manual debugging effort and also decreases the number of iterative verifications.
\end{abstract}

\section{Introduction}\label{sec:intro}
Model checking is a powerful and efficient method for finding flaws in hardware designs,
business processes, object-oriented software, and hypermedia applications.
One remaining major obstacle to a broader application of model checking is its
limited usability for non-experts. In the case of specification violation, it requires much effort and insight to determine the root cause of errors
from the counterexamples generated by model checkers \cite{WNF2010}.

{\sc Web-TLR} \cite{ABR09} is a software tool designed for model-checking
Web applications that is based on rewriting logic~\cite{MM02}. Web applications
are expressed as rewrite theories that can be formally verified by using the Maude built-in LTLR model-checker~\cite{BM08}.
Whenever a property is refuted, a counterexample trace  is delivered   that reveals an undesired, erroneous navigation sequence.
{\sc Web-TLR}  is endowed with support for user interaction in \cite{ABER10}, including the successive exploration of 
error scenarios according to the user's interest by means of a slideshow facility 
that allows the user to incrementally expand
the model states to the desired level of detail, thus avoiding the rather
tedious task of inspecting the textual  representation of the system.
Although this facility helps the user to keep the overview of the model, 
 the analysis (or even the simple inspection) of  the delivered counterexamples 
is still unfeasible because of the size and complexity of the traces under examination.
This is particularly serious   in the rewriting logic context of  {\sc Web-TLR} because
Web specifications may contain equations  and algebraic laws  that are internally used
to simplify the system states, and temporal LTLR formulae may contain function symbols
that are interpreted in the considered algebraic theory.
All of this results in 
execution traces that may be difficult to understand for users who are not acquainted with 
rewriting logic technicalities.

This paper aims at improving the understandability of the  counterexamples generated by   \textsc{Web-TLR}.
This is achieved by means of a complementary Web debugging facility 
that supports both  the efficient manipulation of counterexample traces and the interactive exploration of error scenarios.
This facility is based on a backward trace-slicing technique for rewriting logic theories formalized in~\cite{ABER11} that allows the pieces of information that we are interested to be traced back through the inverse rewrite sequence.
The slicing process  drastically simplifies the computation trace by 
dropping useless data that do not influence the final result.
We provide a convenient, handy notation for specifying the slicing criterion that is successively propagated 
backwards at locations selected by the user.
Preliminary experiments reveal that the novel slicing facility 
of the extended version of {\sc Web-TLR}   is fast enough to enable smooth interaction 
and helps the users to locate the cause of errors accurately without overwhelming them with bulky information.
By using the slicing facility,
the Web engineer can focus on the relevant fragments of the failing   application, which greatly reduces the manual debugging effort.

%

This paper is organized as follows. In Section \ref{sec:example}, we introduce a leading example that allows us  to illustrate the use of rewriting logic
for Web system specification.
An overview of the {\sc Web-TLR} verification framework is given in Section~\ref{sec:webtlr}, and its main facilities are described  in Section~\ref{sec:webtlr-system}.
Section~\ref{sec:imple} 
reports on the extended implementation of the  {\sc Web-TLR}  system.
In   Section~\ref{sec:debug}, we illustrate our methodology for interactive analysis of counterexample traces
and debugging of Web Applications.
  Section~\ref{sec:conclu} concludes.

\section{A Running Example: the Electronic Forum Application}\label{sec:example}

Throughout this paper, a Web application is thought of as a collection of related Web pages that are  
hosted by a Web server
and  contain a mixture of (X)HTML code and executable code (Web scripts), and links to other Web  pages.
A Web application is accessed  over a network such as the Internet  by using 
a Web browser that allows Web pages to be navigated by clicking and following links. Interactions between Web browsers and the Web server are driven by the HTTP protocol. 

As a running example that allows us to illustrate the capabilities of our tool, let us introduce  a Web application that implements an  electronic forum.
The electronic forum  is a rather complex Web system that is equipped with a number of common features, such as user registration, role-based access control including moderator and administrator roles, and topic and comment management.

\begin{figure}[t!]
\centering
\includegraphics[scale=0.45]{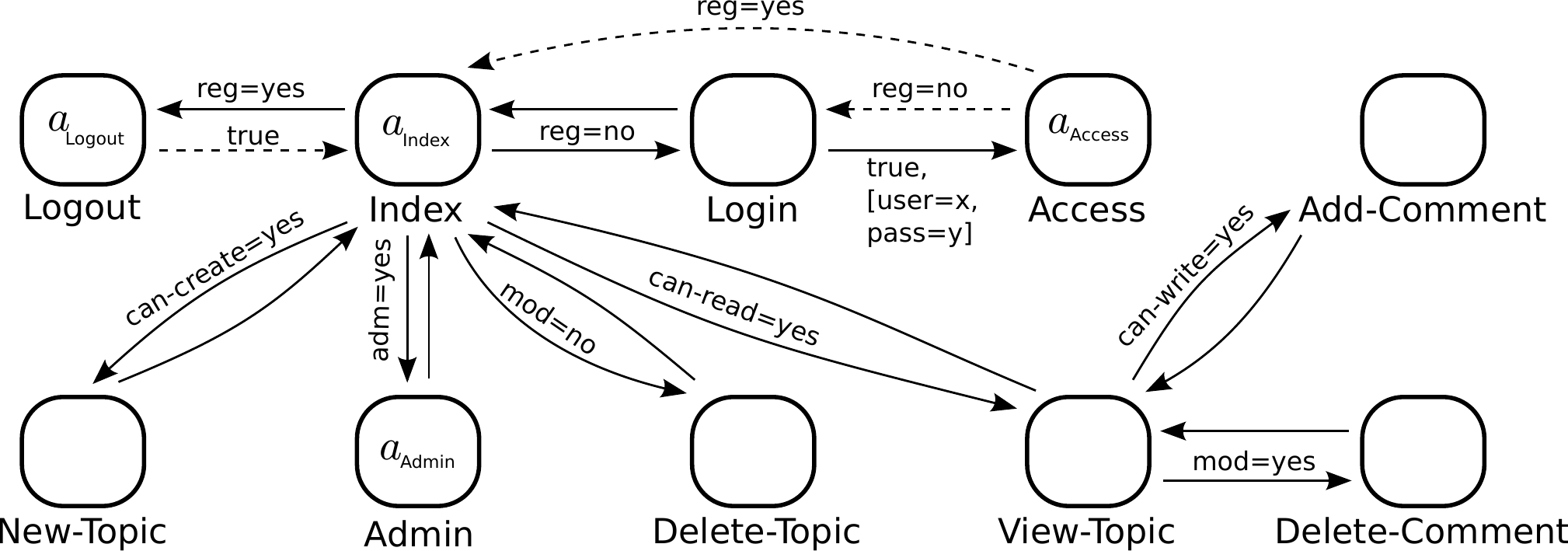}
\caption{The navigation model of an Electronic Forum}
\label{fig:graph}
\end{figure}

Informally speaking, the navigation model (i.e., the intended semantics) of such an application can be specified by means of the graph-like structure given in Figure~\ref{fig:graph}.
Web pages are modeled as graph nodes. Each navigation link~$l$  is specified by a solid arrow that is labeled by a condition $c$ and a query string $q$. Link $l$ is enabled whenever $c$ evaluates to $true$\footnote{Note that conditional links provide a simple but effective form of access control: a Web page can be accessed through a conditional link iff its condition holds.}, while $q$ represents the input parameters that are sent to the Web server once the link is clicked.  
For example, the navigation link that connects the {\sf Login} and {\sf Access} Web pages  is always enabled and requires two input parameters ({\sf user} and {\sf pass}).
The dashed arrows model Web application continuations, that is, arrows pointing to Web pages that are automatically computed by Web script executions. Each dashed arrow is labeled by a condition, which is used to select the continuation at runtime.
For example, the {\sf Access} Web page has got two possible continuations
(dashed arrows) whose labels are {\sf reg=yes} and {\sf reg=no}, respectively. 
The former continuation specifies that  
the login attempt succeeds, and, thus, the {\sf Index} Web page is delivered to the browser; 
in the latter case, the login fails and the {\sf Login} page is sent back to the browser.


\section{An Overview of the RWL-based Web Verification Framework}\label{sec:webtlr}

In this section, we briefly recall the main   features of the Web verification framework proposed in~\cite{ABR09}, which are  essential for  understanding this work. 

\subsection{Web Application Specification}
In our setting, Web applications are specified  by means of a rewrite theory, which accurately formalizes the entities in play (e.g., Web server, Web browsers, Web scripts, Web pages, messages) by means of a rich Web state data structure that 
can be interpreted as a snapshot of the system that captures the current configurations of the active browsers (i.e., the browsers currently using the Web application), together with the server and the channel through which the browsers and the server communicate 
via message-passing.  A part of the Maude specification that formalizes the internal representation of a generic Web state is given in Figure \ref{fig:signature}  (see \cite{ABR09} for a complete specification). 
Formally, a Web state \textsf{WebState} is a triple \textsf{[Browsers] [Messages] [Server]} where  \textsf{Browsers}, \textsf{Messages}, and  \textsf{Server} are encoded by using suitable Maude constructors. For instance, \textsf{Browsers} is a multiset of active browsers that is built using the associative and commutative constructor \textsf{\_;\_}. Each active browser is in turn formalized by a constructor term of the form  $$\sf B(id_b, id_t, page, urls,session, sigma, lm, h, i)$$ where
 ${\sf id_b}$ is an identifier representing the browser; 
 ${\sf id_t}$ is an identifier modeling  an open windows/tab of  browser ${\sf id_b}$;
$\sf page$ is the name of the Web page that is currently displayed on the Web browser; ${\sf urls}$ represents 
the  list of navigation links that appear in the Web page $\sf n$;
 $\sf session$  is  a list of  pairs of the form $(n,v)$ that is used to encode the last session that the server has sent to the browser, where each $n$ represents a server-side variable
whose value is $v$; $\sf sigma$ contains the information that is needed to automatically fill in the forms in the Web pages;
 ${\sf lm}$ is the last message sent to the server; 
 ${\sf h}$ is a bidirectional list that records the history of the visited Web pages;
and ${\sf i}$ is an internal counter used  to distinguish among several response  messages generated by repeated refresh actions (e.g., if a user pressed  the refresh button twice, only the second refresh is displayed in the browser window).

A formal description of the Web pages is encoded  in the \textsf{Server} data structure, together with  the Web application's scripts.
It is worth noting that our scripting language  includes the main features of the most popular Web programming languages such as  built-in primitives  for reading/writing session data ({\sf getSession, setSession}),   accessing and updating data bases ({\sf selectDB, updateDB}), capturing values contained in the query strings sent by  browsers ({\sf getQuery}),  \emph{etc}. 
Figure~\ref{fig:specForum} details the navigation model of the electronic forum application and the Web scripts involved.

The Web application behaviour is formalized by using labeled rewrite rules of the form
$\sf label: WebState \Rightarrow WebState$ that model the application's navigation model through suitable state transitions. 
More specifically, we provide a formal, rule-based description of
\begin{itemize}
\item a request/response, communication protocol that abstracts the main features of HTTP. Our abstraction is equipped with a (sort of) GET method for passing user-defined parameters to the Web server; 
\item an engine for the execution of Web scripts.
\end{itemize}  
For instance, the rule {\sf Evl} shown below consumes the first request message $\sf m_{idb, idt}$ of the queue $\sf fifo_{req}$, evaluates the message w.r.t.\ the corresponding browser session $\sf (id_b, \{s\})$, and generates the response message that is stored in the queue $\sf fifo_{res}$, that is, the server queue that contains the responses to be sent to the browsers.

$$
\begin{array}{l}
{\sf Evl \colon}
{\sf [br][ m ] [S(w, \{BS\underline{(id_b, \{s\})}, bs\}, \underline{\{db\}, (m_{idb, idt}} , fifo_{req}), fifo_{res}) ] }
 \Rightarrow \\
 \hspace{3.7cm} {\sf [ br ] [ m ] [ S(w, \{BS(id_b,\underline{\{s'\}}), bs\}, \underline{\{db'\}} , fifo_{req}, (fifo_{res}, \underline{m')}) ] } \\
{\sf \hspace{0cm} \mbox{where } (s', db', m') = eval(w, s, db,  m_{idb, idt})   }
\end{array}
$$



\begin{figure}
{\small 
\begin{lstlisting}
 op `[_`]`[_`]`[_`] : Browser Message Server -> WebState [ctor] .

 op B : Id Id Qid URL Session Sigma Message History Nat -> Browser [ctor] .
 op br-empty : -> Browser [ctor] .
 op _:_ : Browser Browser -> Browser [ctor assoc comm id:br-empty] .

 op B2S : Id Id URL -> Message [ctor] .               --- message to server
 op S2B : Id Id Qid URL Session -> Message [ctor] .   --- message to browser
 op mes-empty : -> Message [ctor] .
 op _:_ : Message Message -> Message [ctor assoc comm id:mes-empty] .

 op S : Page UserSession DB  ->  Server [ctor] .

 op `(_`,_`,`{_`}`,`{_`}`) : Qid Script Continuation Navigation -> Page [ctor] . 
 op page-empty : -> Page .
 op _:_ : Page Page -> Page [ctor assoc comm id:page-empty] .

\end{lstlisting}
}
\caption{Maude representation of a Web state.}\label{fig:signature}
\end{figure}


\begin{figure}[t!]
Formal description of the navigation model of the electronic forum: 
{\scriptsize
$$\begin{array}{|rl|}
\hline
\sf P_{Index} = & \sf  (Index, \alpha_{index},  \{\emptyset\},  \{  (reg=no) \rightarrow ( Login ? [\emptyset] ) 
                    : (reg=yes) \rightarrow ( Logout ? [\emptyset] )  
                    : (adm=yes) \rightarrow ( Admin ? [\emptyset] ) \\
                & \sf : (can\mbox{-}read=yes) \rightarrow ( View\mbox{-}Topic ? [topic] ) 
                      : (can\mbox{-}create=yes) \rightarrow ( New\mbox{-}Topic ? [topic] ) ) \\
                & \sf : (mod=yes) \rightarrow ( Del\mbox{-}Topic ? [topic] ) \}   )\\
\sf P_{Login} = & \sf ( Login,
                   {\tt skip},
                   \{ \emptyset \},
                   \{ ( \emptyset \rightarrow ( Index ? [\emptyset] ) )
                   : ( \emptyset \rightarrow ( Access ? [user, pass] ) ) \}
                 )\\
\sf P_{Access} = & \sf ( Access,
                    \alpha_{accessScript},
                    \{ ( (reg=yes) \Rightarrow Index )
                    : ( (reg=no) \Rightarrow Login ) \},
                    \{ \emptyset \}
                  )\\
\sf P_{Logout} = & \sf ( Logout,
                    \alpha_{logout},
                    \{ ( \emptyset \Rightarrow Index ) \},
                    \{ \emptyset \} 
                  )\\
\sf P_{Admin} = & \sf ( Admin,
                   \alpha_{admin},
                   \{\emptyset\},
                   \{ ( \emptyset \rightarrow ( Index ? [\emptyset] ) ) \}
                 )\\
\sf P_{AddComment} = & \sf ( AddComment,
                        {\tt skip},
                        \{\emptyset\},
                        \{ ( \emptyset \rightarrow ViewTopic  ? [\emptyset]) \}
                      )\\
\sf P_{DelComment} = & \sf ( DelComment,
                        {\tt skip},
                        \{\emptyset\},
                        \{ ( \emptyset \rightarrow ViewTopic ? [\emptyset] ) \}
                      )\\
\sf P_{ViewTopic} = & \sf ( ViewTopic,
                       {\tt skip},
                       \{\emptyset\},
                       \{ ( \emptyset \rightarrow ( Index ? [\emptyset] ) ) 
                       : ( (can\mbox{-}write=yes) \rightarrow ( AddComment ? [\emptyset] ) ) \\
                    &   \sf : ( (mod=yes) \rightarrow (DelComment ? [\emptyset]) ) \}
                     )\\
\sf P_{NewTopic} = & \sf ( NewTopic,
                      {\tt skip},
                      \{\emptyset\},
                      \{ ( \emptyset \rightarrow ViewTopic ? [\emptyset] ) \}
                    )\\
\sf P_{DelTopic} = & \sf ( DelTopic,
                      {\tt skip},
                      \{\emptyset\}, 
                      \{ ( \emptyset \rightarrow Index ? [\emptyset]) \}
                    )\\
\hline
\end{array}$$
} 


{\noindent Electronic forum Web scripts:}%
\vspace{-.3cm}
\begin{multicols}{2}
\scripting{$\alpha_{\sf access}$}
{   setSession("adm","no"); \\
    setSession("mod", "no") ;\\
    setSession("reg", "no") ;\\
    'u := getQuery('user) ;\\
    'p := getQuery('pass) ;\\
    'p1 := selectDB('u) ;\\
    'createlvl := selectDB("create-level") ;\\
    'writelvl := selectDB("write-level") ;\\
    'readlvl := selectDB("read-level") ;\\
    if ('p = 'p1) then\\
    ~~setSession("user", 'u) ;\\
    ~~'r := selectDB('u '. "-role") ;\\
    ~~setSession("reg", "yes") ;\\
    ~~if ('createlvl = "reg") then\\
    ~~~~setSession("can-create", "yes") fi ;\\
    ~~if ('writelvl = "reg") then\\
    ~~~~setSession("can-write", "yes") fi ;\\
    ~~if ('readlvl = "reg") then\\
    ~~~~setSession("can-read", "yes") fi ;\\
    ~~if ('r = "adm") then\\
    ~~~~setSession("adm" , "yes") ;\\
    ~~~~setSession("mod" , "yes") ;\\
    ~~~~setSession("can-create", "yes") ;\\
    ~~~~setSession("can-write", "yes") ;\\
    ~~~~setSession("can-read", "yes")\\
    ~~else\\
    ~~~~setSession("adm" , "no") ;\\
    ~~~~if ('r = "mod") then\\
    ~~~~~~setSession("mod", "yes") ;\\
    ~~~~~~if ('createlvl = "mod") then\\
    ~~~~~~~~setSession("can-create", "yes") fi ;\\
    ~~~~~~if ('writelvl = "mod") then\\
    ~~~~~~~~setSession("can-write", "yes") fi ;\\
    ~~~~~~if ('readlvl = "mod") then\\
    ~~~~~~~~setSession("can-read", "yes") fi\\
    ~~~~else\\
    ~~~~~~setSession("mod", "no")\\
    fi fi fi}

\scripting{$\alpha_{\sf index}$}
{   
    setSession("adminPage", "free") ;\\
    --- Set default levels\\
    'r := getSession("reg") ;\\
    if ('r = null) then\\
    ~~setSession("reg", "no") ;\\
    ~~setSession("mod", "no") ;\\
    ~~setSession("adm", "no") ;\\
    ~~setSession("can-create", "no") ;\\
    ~~setSession("can-write", "no") ;\\
    ~~setSession("can-read", "no")\\
    fi ;\\
    --- Set capabilities available\\
    'createlvl := selectDB("create-level") ;\\
    'writelvl := selectDB("write-level") ;\\
    'readlvl := selectDB("read-level") ;\\
    if ('createlvl = "all") then\\
    ~~setSession("can-create", "yes")\\
    fi ;\\
    if ('writelvl = "all") then\\
    ~~setSession( "can-write","yes")\\
    fi ;\\
    if ('readlvl = "all") then\\
    ~~setSession("can-read", "yes")\\
    fi
}

\scripting{$\alpha_{\sf logout}$}
{
    setSession("reg", "no") ;\\
    setSession("mod", "no") ;\\
    setSession("adm", "no") ;\\
    setSession("can-create", "no") ;\\
    setSession("can-write", "no") ;\\
    setSession("can-read", "no")
}

\scripting{$\alpha_{\sf admin}$}
{
   setSession("adminPage", "busy")
}
\end{multicols}
\caption{Specification of the electronic forum application in {\sc Web-TLR}}\label{fig:specForum}
\end{figure}

\subsection{Model-checking Web Applications}
Formal properties of the Web application 
can be specified by means of the {\em Linear Temporal Logic of Rewriting} (LTLR), which is a temporal logic that extends the traditional Linear Temporal Logic (LTL) with \emph{state predicates} \cite{Meseguer2008},
i.e, 
atomic predicates that are locally evaluated on the states of the system.
Let us see some examples.
Consider again the electronic forum example of Section \ref{sec:example} along with the Maude code in Figure \ref{fig:signature} that describes our Web state structure.
We can define the state predicate ${\sf curPage(id_b, page)}$ by means of a boolean-value function as follows, 
$$\sf  [B(\underline{id_b}, id_t, \underline{page} , urls, session, sigma, lm, h, i), br ][ m ][ sv ] 
   \sf \models curPage(id_b, page) = true$$
which holds (i.e., evaluates to {\sf true}) for any Web state such that {\sf page} is the current Web page  displayed in the browser with identifier ${\sf id_b}$.

%
%
%

By defining elementary state predicates, we can build more complex LTLR formulas that express mixed properties containing dependencies among states, actions, and time. 
These properties intrinsically involve  both action-based and state-based aspects  that are either not expressible or are difficult to express in other temporal logic frameworks.
For example, consider the administration Web page {\sf Admin} of the electronic forum application. 
Let us consider two administrator users whose identifiers are {\sf bidAlfred} and {\sf bidAnna}, respectively. 
Then, the mutual exclusion property {\em ``no two administrators can access the administration page simultaneously''} can be defined as follows.
\begin{equation}
\sf \square \neg (curPage(bidAlfred, Admin) \wedge curPage(bidAnna, Admin))\label{property}
\end{equation}

Any given LTLR property can be automatically checked by using the built-in LTLR model-checker \cite{Meseguer2008}. If the property of interest is not satisfied, a counter-example that consists of the erroneous trace is returned.
This trace is expressed as a sequence of rewrite steps that leads from the initial state to the state that violates the property.
Unfortunately, the analysis (or even the simple inspection) of these traces may be unfeasible because of the size and complexity of the traces under examination. 
Typical counter-example traces in {\sc Web-TLR} consist in a sequence of around 100 states, each of which contains more than 5.000 characters.
As an example,   one of the Web states that corresponds to the example given in Section~\ref{sec:debug} is shown in Figure~\ref{fig:counterexample}, which demonstrates  that the manual analysis of counterexample traces is generally impracticable  for debugging purposes.

\begin{figure}[t!]
{\tiny \sf
\renewcommand{\baselinestretch}{.5} 
\{[ B(\underline{\bf \scriptsize bidAlfred}, tidAlfred, \underline{\bf \scriptsize 'Admin}, 'Index ? query-empty, (s("adm"), s("yes"))  :  (s("adminPage"), s("busy"))  :  (s("can-create"), s("yes"))  :  (s("can-read"), s("yes"))  :  (s("can-write"), s("yes"))  :  (s("mod"), s("yes"))  :  (s("reg"), s("yes")), ('pass / "secretAlfred")  :  'user / "alfred", m(bidAlfred, tidAlfred, 'Admin ? query-empty, 1), history-empty, 1)  :  B(\underline{\bf \scriptsize bidAnna}, tidAnna, \underline{\bf \scriptsize 'Admin}, 'Index ? query-empty, (s("adm"), s("yes"))  :  (s("adminPage"), s("busy"))  :  (s("can-create"), s("yes"))  :  (s("can-read"), s("yes"))  :  (s("can-write"), s("yes"))  :  (s("mod"), s("yes"))  :  (s("reg"), s("yes")), ('pass / "secretAnna")  :  'user / "anna", m(bidAnna, tidAnna, 'Admin ? query-empty, 1), history-empty, 1)]bra-empty[mes-empty][S(('Access, setSession(s("adm"), s(    "no"));  setSession(s("mod"), s("no"));  setSession(s("reg"), s("no"));  'u  :=     getQuery('user);  'p  :=  getQuery('pass);  'p1  :=  selectDB('u);  'createlvl  :=     selectDB(s("create-level"));  'writelvl  :=  selectDB(s("write-level"));     'readlvl  :=  selectDB(s("read-level"));  if 'p = 'p1  then  'r  :=  selectDB('u    '. s("-role"));  setSession(s("reg"), s("yes"));  if 'createlvl = s("reg") then     setSession(s("can-create"), s("yes"))fi ;  if 'writelvl = s("reg") then     setSession(s("can-write"), s("yes"))fi ;  if 'readlvl = s("reg") then     setSession(s("can-read"), s("yes"))fi ;  if 'r = s("adm") then  setSession(s(    "adm"), s("yes"));  setSession(s("mod"), s("yes"));  setSession(s(    "can-create"), s("yes"));  setSession(s("can-write"), s("yes"));  setSession(s(    "can-read"), s("yes"))else setSession(s("adm"), s("no"));  if 'r = s(    "mod") then  setSession(s("mod"), s("yes"));  if 'createlvl = s("mod") then     setSession(s("can-create"), s("yes"))fi ;  if 'writelvl = s("mod") then     setSession(s("can-write"), s("yes"))fi ;  if 'readlvl = s("mod") then     setSession(s("can-read"), s("yes"))fi else setSession(s("mod"), s("no"))fi fi    fi, \{(s("reg") '==  s("no")  $=>$  
 'Login) : (s("reg") '==  s("yes")  $=>$   'Index)\}, \{    nav-empty\}) : ('Add-Comment, skip, \{cont-empty\}, \{(TRUE   $->$   'View-Topic  ?     query-empty)\}) : ('Admin, setSession(s("adminPage"), s("busy")), \{cont-empty\}, \{(    TRUE   $->$   'Index  ?  query-empty)\}) : ('Delete-Comment, skip, \{cont-empty\}, \{(TRUE      $->$   'View-Topic  ?  query-empty)\}) : ('Delete-Topic, skip, \{cont-empty\}, \{(TRUE   $->$      'Index  ?  query-empty)\}) : ('Index, setSession(s("adminPage"), s("free"));  'r  :=     getSession(s("reg"));  if 'r = null  then  setSession(s("reg"), s("no"));     setSession(s("mod"), s("no"));  setSession(s("adm"), s("no"));  setSession(s(    "can-create"), s("no"));  setSession(s("can-write"), s("no"));  setSession(s(    "can-read"), s("no"))fi ;  'createlvl  :=  selectDB(s("create-level"));     'writelvl  :=  selectDB(s("write-level"));  'readlvl  :=  selectDB(s(    "read-level"));  if 'createlvl = s("all") then  setSession(s("can-create"), s(    "yes"))fi ;  if 'writelvl = s("all") then  setSession(s("can-write"), s(    "yes"))fi ;  if 'readlvl = s("all") then  setSession(s("can-read"), s(    "yes"))fi, \{cont-empty\}, \{(s("adm") '==  s("yes")  $->$   'Admin  ?  query-empty) : (s(    "can-create") '==  s("yes")  $->$   'New-Topic  ?  'topic '= "") : (s("can-read") '==  s(    "yes")  $->$   'View-Topic  ?  'topic '= "") : (s("mod") '==  s("yes")  $->$   'Delete-Topic     ?  'topic '= "") : (s("reg") '==  s("no")  $->$   'Login  ?  query-empty) : (s("reg") '==     s("yes")  $->$   'Logout  ?  query-empty)\}) : ('Login, skip, \{cont-empty\}, \{(TRUE   $->$      'Access  ? ('pass '= "") :  'user '= "") : (TRUE   $->$   'Index  ?  query-empty)\}) : (    'Logout, setSession(s("reg"), s("no"));  setSession(s("mod"), s("no"));     setSession(s("adm"), s("no"));  setSession(s("can-create"), s("no"));     setSession(s("can-write"), s("no"));  setSession(s("can-read"), s("no")), \{(    TRUE   $=>$   'Index)\}, \{nav-empty\}) : ('New-Topic, skip, \{cont-empty\}, \{(TRUE   $->$      'View-Topic  ?  query-empty)\}) : ('View-Topic, skip, \{cont-empty\}, \{(TRUE   $->$      'Index  ?  query-empty) : (s("can-write") '==  s("yes")  $->$   'Add-Comment  ?     query-empty) : (s("mod") '==  s("yes")  $->$   'Delete-Comment  ?  query-empty)\}), us(bidAlfred, (s("adm"), s("yes"))  :  (s("adminPage"), s("busy"))  :  (s("can-create"), s("yes"))  :  (s("can-read"), s("yes"))  :  (s("can-write"), s("yes"))  :  (s("mod"), s("yes"))  :  (s("reg"), s("yes")))  :  us(bidAnna, (s("adm"), s("yes"))  :  (s("adminPage"), s("busy"))  :  (s("can-create"), s("yes"))  :  (s("can-read"), s("yes"))  :  (s("can-write"), s("yes"))  :  (s("mod"), s("yes"))  :  (s("reg"), s("yes"))), mes-empty, readymes-empty, (s("alfred") ;  s("secretAlfred")) (s("alfred-role") ;  s("adm")) (s("anna") ;  s("secretAnna")) (s("anna-role") ;  s("adm")) (s("create-level") ;  s("reg")) (s("marc") ;  s("secretMarc")) (s("marc-role") ;  s("mod")) (s("maude") ;  s("secretMaude")) (s("maude-role") ;  s("mod")) (s("rachel") ;  s("secretRachel")) (s("rachel-role") ;  s("reg")) (s("read-level") ;  s("all")) (s("robert") ;  s("secretRobert")) (s("robert-role") ;  s("reg")) (s("write-level") ;  s("reg")))] ,  'ReqFin\}

}
\caption{One Web state of the counter-example trace of Section~\ref{sec:debug}.} \label{fig:counterexample}
\end{figure}

\section{Extending the {\sc Web-TLR} System}\label{sec:webtlr-system}

{\sc Web-TLR} is a model-checking tool that implements the theoretical framework of~\cite{ABR09}.
The \textsc{Web-TLR} system is available online 
via its friendly Web interface  at \url{http://www.dsic.upv.es/~dromero/web-tlr.html}.
The Web interface  frees users from having to install applications on their local computer and hides unnecessary technical details of the tool operation. 
After introducing  the 
(or customizing a default)   Maude specification of a Web application, together with  an initial Web state 
 ${st}_0$ and the LTLR formula $\varphi$ to be verified,   $\varphi$ can be automatically checked  at  ${st}_0$.
Once all 
inputs have been entered in the system, we can automatically check the property
 by just clicking the button {\sf Check}, which invokes the Maude built-in operator {\sf tlr~check}\cite{BM08} that supports model checking of LTLR formulas in rewrite theories.
If the property is not satisfied, an interactive slideshow that illustrates the corresponding counterexample (expressed in the form of an execution trace) is generated.
The slideshow supports both forward and backward navigation through the execution trace and combines a graphical representation of the application's navigation model with a detailed textual description of the Web states.

Although Web-TLR provides a complete picture of both, the application model and the generated counterexample, this information is hardly exploitable for debugging Web applications. Actually, the graphical representation provides a very coarse-grained model of the application's dynamics, while the textual description 
conveys too much information (e.g., see Figure~\ref{fig:counterexample}). Therefore, in several cases both representations may result in limited use. 

In order to assist Web engineers in the debugging task, we extend {\sc Web-TLR} by including a trace-slicing technique whose aim is to reduce the amount of information recorded by the textual description of counterexamples. 
Roughly speaking, this technique (originally described in \cite{ABER11}) consists in tracing back, along an execution trace, all the symbols of a (Web) state that are of interest (target symbols), while useless data are discarded.
The basic idea is to take a Rewriting Logic execution trace and traverse  it backwards in order to filter out data that are definitely related to the wrong behavior. This way, we can focus our attention on the most critical parts of the trace, which are eventually responsible for the erroneous application's behaviour.  
It is worth noting that our trace slicing procedure is sound in the sense that, given an execution trace $\cT$,
it automatically computes a trace slice of $\cT$ that includes all the information needed to produce the target symbols of $\cT$ 
we want to observe. In other words, there is no risk that our tool eliminates data from the original execution trace $\cT$ which are indeed relevant w.r.t.\ the considered target symbols.  
Soundness of backward trace slicing has been formally proven in \cite{ABER11tr}. 
  

We have implemented the backward
trace-slicing technique as a stand-alone application written in Maude that can be used to simplify  general Maude traces (e.g., the
ones printed when the trace is set on in a standard rewrite). Furthermore,
we have coupled   the on-line {\sc Web-TLR} system  
with  the slicing tool  in order to optimize the counterexample traces delivered by  {\sc Web-TLR}. 
To achieve this, the external slicing routine is fed with the given counterexample, the selected Web state $s$ 
where the backward-slicing process is required to start, and the slicing criterion for $s$ ---that is, the symbols of $s$ we want to trace back.
It is worth noting that, for model checking Web applications with Web-TLR, we have developed a specially--tailored, handy filtering notation that allows us to easily specify  the slicing criterion and automatically select the desired information 
by exploiting the powerful, built-in  pattern-matching mechanism of Rewriting Logic. 
%
The outcome of the slicing process is a sliced version of the textual description of the original counterexample trace 
which  facilitates the interactive exploration of error scenarios when debugging Web applications.  

%

\subsection{Filtering Notation}\label{sec:wildcard}


In order to select the relevant information to be traced back, we introduce  a simple, pattern-matching filtering language 
that
frees the user from explicitly introducing
the specific positions of the Web state that s/he wants to observe~\footnote{Terms are viewed as labeled trees in the usual way. 
Positions are represented by sequences of natural numbers denoting an access path in a term. 
The empty sequence $\Lambda$ denotes the root position of the term.}. 
Roughly speaking, the user introduces an information pattern $p$ 
that has to be detected inside a given Web state $s$.
The information matching $p$ that is recognized in $s$, is then identified by pattern matching and is kept in $s^\bullet$, whereas 
all other symbols of $s$ are considered irrelevant and then removed.
Finally, the positions of the Web state where the relevant information is located are obtained from $s^\bullet$.
In other words, the slicing criterion is defined by the set of positions where the  relevant information is located within the state $s$ that we are observing and is automatically generated by pattern-matching the  information pattern against the Web state $s$.

The filtering language allows us to define the relevant information as follows:
$(i)$  by giving the name of an operator (or constructor)  or a substring of it; and
$(ii)$ by using the question mark ``$\sf ?$'' as a wildcard character that indicates the position where the information is considered relevant. 
On the other hand, the irrelevant information can be declared by using the wildcard symbol ``$\sf \_$'' as a placeholder for uninteresting arguments of an operator.

Let us illustrate this filtering notation by means of a rather intuitive example.
%
%
Let us assume that the electronic forum application
allows   one to list some data about the available topics. Specifically, the following term $t$ 
specifies the names of the topics available in our electronic forum together with the total number of
posted messages  for each topic.
{\small $$\begin{array}{l}
\sf topic\_info(
topic(astronomy,\sharp posts(520)),
topic(stars,\sharp posts(58)), \\
\sf \hspace*{1.55cm}
topic(astrology,\sharp posts(20)),
topic(telescopes,\sharp posts(290)) ~
)
\end{array}
$$}%

Then, the pattern $\sf topic(astro,\sharp posts(?))$ defines a slicing criterion that allows us to observe the  topic name as well as the total number of messages 
for all  topics whose name includes the word {\sf astro}. 
Specifically, by applying such a pattern to the term $t$, we obtain the following term slice 
{\small $$
\begin{array}{l}
\sf topic\_info(
topic(astronomy,\sharp posts(520)),
\bullet, 
topic(astrology,\sharp posts(20)),
\bullet)
\end{array}
$$}%
which ignores the information related to the topics {\sf stars} and {\sf telescopes},  
and  induces the slicing criterion 
$$\{
\Lambda.1.1,~
\Lambda.1.2.1,~
\Lambda.3.1,~
\Lambda.3.2.1\}.$$

Note that we have introduced the fresh symbol $\bullet$ to approximate any output information in the term that is not relevant with respect to a given pattern.

\section{Implementation of the extended \textsc{Web-TLR} system in RWL}\label{sec:imple}

The enhanced verification 
methodology described in this paper has been implemented in the \textsc{Web-TLR} system using the high-performance, rewriting logic language Maude~\cite{maude-book}.
In this section, we discuss some of the most important features of the Maude language that 
we have been conveniently exploited for the optimized implementation of {\sc Web-TLR}. 

Maude is a high-performance, reflective language that supports both equational and rewriting logic programming, which is particularly suitable for developing domain-specific applications \cite{EMS03,EMM06}. 
In addition, the Maude language is not only intended for system prototyping, but it has to be considered as a real programming language with competitive performance.
The   salient features of Maude  that we used in the implementation of our framework are as follows.

\subsection*{Metaprogramming}

Maude is based on rewriting logic \cite{MM02}, which is reflective in a precise mathematical way.
In other words, there is a finitely presented rewrite theory 
$\cU$ that is universal in the sense that we can represent 
any finitely presented rewrite theory $\cR$  (including $\cU$ itself) in $\cU$ (as a datum), 
and then mimick the behavior of~$\cR$ in $\cU$. 

In the implementation of the extended {\sc Web-TLR} system, we have   exploited the metaprogramming capabilities of Maude in order to provide the system with our backward-tracing slicing tool for RWL theories in RWL itself.
Specifically, during the backward-tracing slicing  process, all input {\sc Web-TLR} modules are   raised to the meta-level and handled as meta-terms, which are meta-reduced and meta-matched by Maude operators.

\subsection*{AC Pattern Matching}
The evaluation mechanism of Maude is based on rewriting modulo an equational theory $E$ (i.e., a set of equational axioms), which is  accomplished by performing {\em pattern matching modulo} the equational theory $E$. More precisely, given an equational theory {\em E}, a term {\em t} and a  term {\em u}, we say that {\em t matches u modulo E} (or that {\em t E-matches u}) if there is a substitution $\sigma$ such that {\em $\sigma$(t)$=_E$ u}, that is, {\em $\sigma$(t)} and {\em u} are equal modulo the equational theory {\em E}.
When $E$ contains axioms that express the associativity and commutativity of one operator, we talk about {\em AC pattern matching}. 
We have exploited the AC pattern matching 
to implement  both the filtering language 
and   the slicing process.

\subsection*{Equational Attributes}
Equational attributes are a means of declaring certain kinds of equational axioms in a way that allows Maude to use these equations efficiently in a built-in way.
Semantically, declaring a set of equational attributes for an operator is equivalent to declaring the corresponding equations for the operator. 
In fact, the effect of declaring equational attributes is to compute with equivalence classes modulo these equations. 
This avoids termination problems and leads to much more efficient evaluation.

In the signature presented in Figure~\ref{fig:signature}, the overloaded operator {\sf \_:\_} is given with the equational attributes {\sf assoc}, {\sf comm}, and {\sf id}.
This  allows Maude to handle simple objects and multisets of elements in the same way.
For example, given two terms $\sf b_1$ and $\sf b_2$ of sort {\sf Browser}, the term  $\sf b_1 : b_2$ belongs to the sort {\sf Browser} as well.
Also, these equational attributes allow us to get rid of parentheses and disregard the ordering among elements.
For example, the communication channel is modeled as a term of sort {\sf Message} where the messages among the browsers and the server can arrive out of order, which allows us to simulate the HTTP communication protocol.

\subsection*{Flat/unflat Transformations}

In Maude, AC pattern matching is implemented by means of a special encoding of AC operators, which allows us to represent 
AC terms terms by means of  single representatives that are obtained by replacing nested occurrences of the same AC operator by a flattened argument list under a variadic symbol, whose elements are sorted by means of some linear ordering\footnote{Specifically, Maude uses the lexicographic order of symbols.}.
The inverse of the flat transformation is the unflat transformation, which is nondeterministic in the sense that it generates all the unflattended terms that are equivalent (modulo AC) to the  flattened term. 
For example, consider a binary AC operator $f$  together with the standard lexicographic ordering over symbols. 
Given the $AC$-equivalence $f(b,f(f(b,a),c)) =_{AC} f(f(b,c),f(a,b))$, we can represent it by using the ``internal sequence'' $f(b,f(f(b,a),c)) \raw^*_{\mathit{flat}_{AC}} f(a,b,b,c) \raw^*_{\mathit{unflat_{AC}}} f(f(b,c),f(a,b))$, where the first subsequence corresponds to the {\em flattening} transformation that obtains the AC canonical form of the term, whereas the second one corresponds to the inverse, unflattening transformation. 

These two processes are typically hidden inside the $AC$-matching algorithms\footnote{See~\cite{maude-manual} (Section~$4.8$) for an in-depth discussion on matching and simplification modulo AC in Maude.} that are used to implement the rewriting modulo relation. 
In order to facilitate the understanding of the sequence of rewrite steps, we exposed the flat and unflat transformations visibly in our slicing process. 
This is done by breaking up a rewrite step and adding the intermediate $\sf flat/unflat$  transformation sequences into the computation trace delivered by Maude.

\section{A Debugging Session with {\sc Web-TLR}}\label{sec:debug}

In this section, we illustrate our methodology for interactive analysis of counterexample traces
and debugging of Web applications.

Let us consider an initial state that consists of two administrator users whose identifiers are {\sf bidAlfred} and {\sf bidAnna}, respectively.
Let us also recall the mutual exclusion Property \ref{property} of Section \ref{sec:webtlr} 
{ $$\sf \square\,\neg\, (curPage(bidAlfred,Admin) \wedge  curPage(bidAnna,Admin))$$}%
which states that ``no two administrators can access the administration page simultaneously''.
Note that the predicate state $\sf curPage(bidAlfred, Admin)$
holds when the user $\sf bidAlfred$ logs into the $\sf Admin$ page (a similar interpretation is given to predicate $\sf curPage(bidAnna,Admin)$).
By verifying the above property with \textsc{Web-TLR}, we get a huge counterexample that proves that the property is not satisfied.
The trace size weighs around $190kb$.

In the following, we show how the considered Web application can be debugged using  {\sc Web-TLR}.
First of all, 
we specify the slicing criterion to be applied on the counterexample trace.
This is done by using the wildcard notation on the terms   introduced in Section \ref{sec:wildcard}.
Then, the slicing process is invoked and the resulting trace slice is produced.
Finally, we analyze the trace slice and 
outline a methodology that helps 
the user to locate the errors.

\subsubsection*{Slicing Criterion}

%

The slicing criterion represents the information that we want to trace back through the execution trace~$\cT$ that is produced as the outcome of the 
{\sc Web-TLR} model-checker. 

For example, consider the final Web state $s$ shown in Figure~\ref{fig:counterexample}.
In this Web state, the two users, {\tt bidAlfred} and {\tt bidAnna}, are logged into the {\tt Admin} page.
Therefore, 
the considered mutual exclusion property has been violated.
Let us assume that we want to diagnose the erroneous pieces of information within the execution trace $\cT$ that produce this bug. 
Then, we can enter the following information pattern as input,
$$\sf B(?,\_,?,\_,\_,\_,\_,\_,\_)$$
where the operator {\sf B} restricts the search of relevant information inside the browser data structures,  
the first question symbol {\sf ?} represents that we are interested in tracing the user identifiers, and the second one calls for the Web page name. 
Thus, by applying the considered information pattern to the Web state $s$, we obtain the slicing criterion $\{ \Lambda.1.1.1, ~\Lambda.1.1.3, ~\Lambda.1.2.1, ~\Lambda.1.2.3 \}$ and the corresponding sliced state 
$$ s^\bullet=\sf  [B(bidAlfred,\bullet,Admin,\bullet,\bullet,\bullet,\bullet,\bullet,\bullet) : B(bidAnna,\bullet,Admin,\bullet,\bullet,\bullet,\bullet,\bullet,\bullet)] [\bullet] [\bullet]  $$ 

Note that  $\Lambda.1.1.1$ and $\Lambda.1.2.1$ are the positions in $s^\bullet$ of the user identifiers  {\tt bidAlfred} and {\tt bidAnna}, respectively, and $\Lambda.1.1.3$ and $\Lambda.1.2.3$ 
are the positions in $s^\bullet$ that indicate that the users are logged into the $\tt Admin$ page.

\subsubsection*{Trace Slice}
Let us consider the counterexample execution trace $\cT=s_0\raw s_1\raw\ldots \raw s_n$, where $s_n=s$. 
The slicing technique proceeds backwards, from the observable state $s_n$ to the initial state $s_0$, and for each state $s_i$ recursively generates a sliced state $s_i^\bullet$ that consists of the relevant information with respect to the slicing criterion.

By running the backward-slicing tool with the execution trace $\cT$ and the slicing criterion 
given above as input, we get the trace slice $\cT^\bullet$ as outcome, where  
 useless data that do not influence the final result are discarded.
Figure~\ref{fig:slice-trace} shows a part of the trace slice $\cT^\bullet$. 

It is worth observing that  the slicing process greatly reduces the size of the original trace $\cT$, and allows us to center on those data that are likely to be the source of an erroneous behavior.  

Let  $\mid \!\cT\!\mid$ be the size of the trace $\cT$, namely the sum of the number of symbols of all trace states.
In this specific case, the size reduction that is achieved on the
the subsequence $s_{(n-6)} \ldots s_{n}$ of $\cT$, in symbols $\cT_{[s_{(n-6)}.. s_{n}]}$ 
is:
$${\mid\!\cT^\bullet_{[s_{(n-6)}^\bullet \ldots {s_n^\bullet}]}\!\mid \over \mid\!\cT_{[s_{(n-6)} \ldots {s_n}]}\!\mid} = {121 \over 1458} = 0.083 \mbox{ (i.e., a reduction of 91.7\%)}
$$

\begin{figure}[t!]
\small
$$
\begin{array}{ll}
\cT^\bullet = s_0^\bullet  
\ldots & \raw
s_{n-6}^\bullet  \overarrow{ScriptEval}
s_{n-5}^\bullet  \overarrow{flat/unflat}
s_{n-4}^\bullet  \overarrow{ResIni}
s_{n-3}^\bullet \\ &  \\ &  \overarrow{flat/unflat}  
s_{n-2}^\bullet \overarrow{ResFin} 
s_{n-1}^\bullet \overarrow{flat/unflat} 
s_{n}^\bullet 
\end{array}
$$


\renewcommand{\baselinestretch}{.5} 
where
{\small

$$
\begin{array}{rl}

\bigskip

s_n^\bullet = & 
\tt [B(\underline{bidAlfred},*, \underline{Admin},*,*,*,*,*,*) : B(bidAnna,*, Admin,*,*,*,*,*,*)]* 
\tt [*]
\tt [*]  \\

\bigskip

s_{n-1}^\bullet = &  
\tt [B(bidAnna,*, Admin,*,*,*,*,*,*) : B(\underline{bidAlfred},*, \underline{Admin},*,*,*,*,*,*)]* 
\tt [*] 
\tt [*] \\

s_{n-2}^\bullet = & 
\tt [B(bidAnna,*, Admin,*,*,*,*,*,*) : B(\underline{bidAlfred}, \underline{tidAlfred},*,*,*,*,*,*, 1)]* \\
\bigskip
 & \tt [\underline{ S2B(bidAlfred,tidAlfred, Admin,*,*, 1)} : *]
 \tt [*] \\

s_{n-3}^\bullet = &
\tt [B(\underline{bidAlfred},\underline{tidAlfred},*,*,*,*,*,*,1) : B(bidAnna,*,Admin,*,*,*,*,*,*)]*  \\ 
\bigskip
 & \tt [* : \underline{ S2B(bidAlfred,tidAlfred, Admin,*,*,1) } ] 
 \tt [*] \\

s_{n-4}^\bullet = & 
\tt [B(bidAlfred, tidAlfred,*,*,*,*,*,*,1) : B(bidAnna,*, Admin,*,*,*,*,*,*)] *
\tt [*]   \\ 
\bigskip
 &  \tt  [S(*, * : us(bidAlfred,*), *, (rm( \underline{ S2B(bidAlfred, tidAlfred , Admin,*,*,1) }, *, *) : *), *] \\

s_{n-5}^\bullet = & 
\tt [B(bidAlfred, tidAlfred,*,*,*,*,*,*,1) : B(bidAnna,*,Admin,*,*,*,*,*,*)] *
\tt [*] \\
 & \tt [S(*, us(bidAlfred,*) : *, *, ( * : 
 \underline{ evalScript }(WEB\mbox{-}APP, SESSION, \\ 
\bigskip
 & \tt \underline{ B2S(bidAlfred, tidAlfred, Admin ? query\mbox{-}empty, 1) } ,DB) ) , * )] \\

s_{n-6}^\bullet = &
\tt [B(bidAlfred, tidAlfred,*,*,*,*,*,*,1) : B(bidAnna,*, Admin,*,*,*,*,*,*)] * [*] \\
& \tt [S(WEB\mbox{-}APP, (* : us(bidAlfred, SESSION)),  \\
 & \tt \underline{ B2S(bidAlfred, tidAlfred, Admin ? query\mbox{-}empty ,1) }: * , *, DB ] 

\end{array}
$$
}

\caption{Trace slice $\cT^\bullet$.}\label{fig:slice-trace}
\end{figure}

\subsubsection*{Trace Slice Analysis}

Let us analyze   the information recorded in the trace slice $\cT^\bullet$.
In order to facilitate  understanding, the main symbols involved in the  description are underlined in Figure~\ref{fig:slice-trace}.

\begin{itemize}

\item [-]
The sliced state $s_{n}^\bullet$ is the observable state that records only the relevant information defined by the slicing criterion. 

\item [-]
The slice state $s_{n-1}^\bullet$ is obtained from $s_n^\bullet$ by the flat/unflat transformation.

\item[-] 
In the sliced state $s_{n-2}^\bullet$, the communication channel contains a response message for the user {\tt bidAlfred}.
This response message enables the user {\tt bidAlfred} to log into the {\tt Admin} page.
Note that the identifier {\tt tidAlfred} occurs in the Web state.
This identifier signals the open window that the response message refers to.
Also, the number {\tt 1} that occurs in the sliced state $s_{n-2}^\bullet$ 
represents the {\em ack} (acknowledgement) of the response message.
Finally, the reduction from $s_{n-2}^\bullet$ to $s_{n-3}^\bullet$ corresponds again to a flat/unflat transformation.

\item[-]
In the sliced state $s_{n-4}^\bullet$, we can see the response message stored in the server that is ready to be sent, whereas, in the server configuration of the sliced state $s_{n-5}^\bullet$, the operator $\tt evalScript$ occurs. 
This operator takes the Web application ({\tt WEB-APP}), the user session ({\tt SESSION}), the request message, and the database ({\tt DB}) as input. 
The request message contains the query string that has been sent by the user {\tt bidAlfred} to ask for admission into the {\tt Admin} page.
Observe that the response message that is shown in the slice state $s_{n-4}^\bullet$ is the one given as the outcome of the evaluation of the operator $\tt evalScript$ in the sliced state $s_{n-5}^\bullet$.

\item[-]
Finally, the sliced state $s_{n-6}^\bullet$ shows the request message waiting to be evaluated.

\end{itemize}

\begin{figure}[h!]
\centering
\fbox{
\includegraphics[width=.87\textwidth]{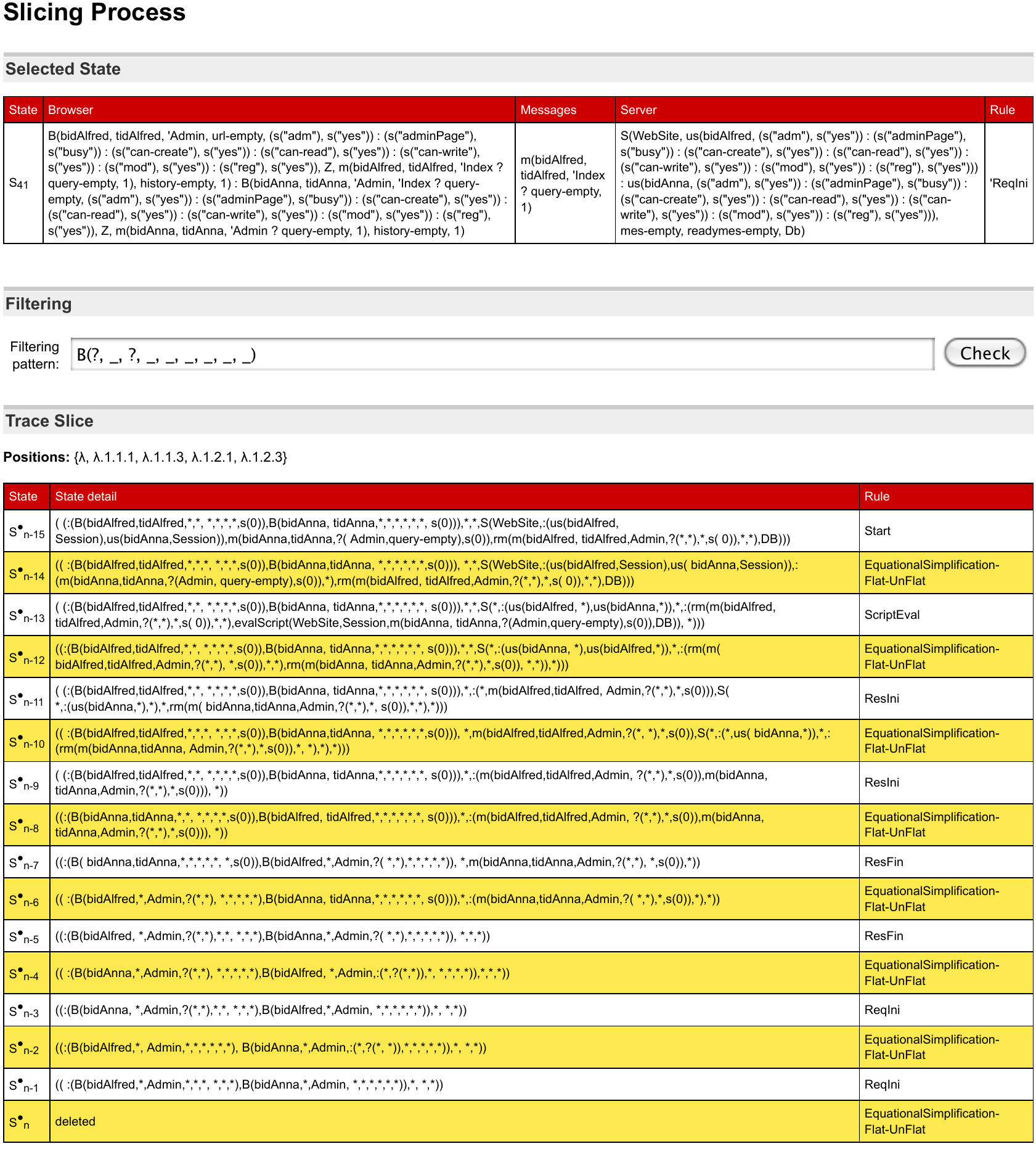}
}
\caption{Snapshot of the {\sc Web-TLR} System.}\label{fig:slicing-process}
\end{figure}

Note that the outcome delivered by the operator {\tt evalScript}, when the script $\alpha_{\sf admin}$ is evaluated, is not what the user would have expected, since it allows the user to log into the {\sf Admin} page, which leads to the violation of the considered property.
This identifies  the script $\alpha_{\sf admin}$ as the script that is responsible for the error. 
Note that this conclusion is correct because $\alpha_{\sf admin}$ has not implemented a mutual exclusion control (see Figure~\ref{fig:specForum}).
A snapshot of {\sc Web-TLR} that shows the slicing process is given in Figure~\ref{fig:slicing-process}.

This bug can be fixed by introducing the necessary control for mutual exclusion as follows.
First, a continuation $\sf ("adminPage" = "busy") \rightarrow  Index ? [\emptyset])$ is added to the {\sf Admin} page, and the $\sf \alpha_{admin}$ is replaced by a new Web script that checks whether there is  another user in the {\sf Admin} page.
In the case when the {\sf Admin} page is {\sf busy} because it is being accessed by a given user, any other user is redirected to the {\sf Index} page.
If the {\sf Admin} page is {\sf free}, the user asking for permission to enter is authorized to do so (and the page gets  locked).
Furthermore, the control for unlocking the {\sf Admin} page is added at the beginning of the script $\sf \alpha_{index}$.
Hence, the fixed Web scripts are  as follows:

%
%
%
%
%


\begin{center}
\fbox{
$\begin{array}{rl}
\sf P_{Admin} = & \sf ( Admin,
                   \alpha_{admin}, \\
                & \sf   \{ \underline{("adminPage" = "busy") \rightarrow  Index ? [\emptyset]) } \},\\
                & \sf   \{ ( \emptyset \rightarrow ( Index ? [\emptyset] ) ) \}
                 )
\end{array}$
}
\end{center}

\begin{multicols}{2}

where the new $\alpha_{\sf Admin}$ is:
\scripting{$\alpha_{\sf admin}$}
{	'u := getSession("user") ; \\
	'adm := selectDB("adminPage") ;\\
	if ( 'adm != 'u) then \\
	~~~setSession("adminPage", "busy") \\
	else \\
	~~~setSession("adminPage", "free")) ;\\
	~~~updateDB( "adminPage", 'u) ) \\
	fi
}

and the piece of code that patches $\alpha_{\sf index}$ is:

\scripting{$\alpha_{\sf index}$}
{	'adm := getSession("adminPage") ; \\
    if ('adm = "free") then \\
	~~~updateDB("adminPage", "free") \\
    fi ; \\
    \ldots   
}
\end{multicols}

\begin{figure}[h!]
\centering
\fbox{
\includegraphics[width=.85\textwidth]{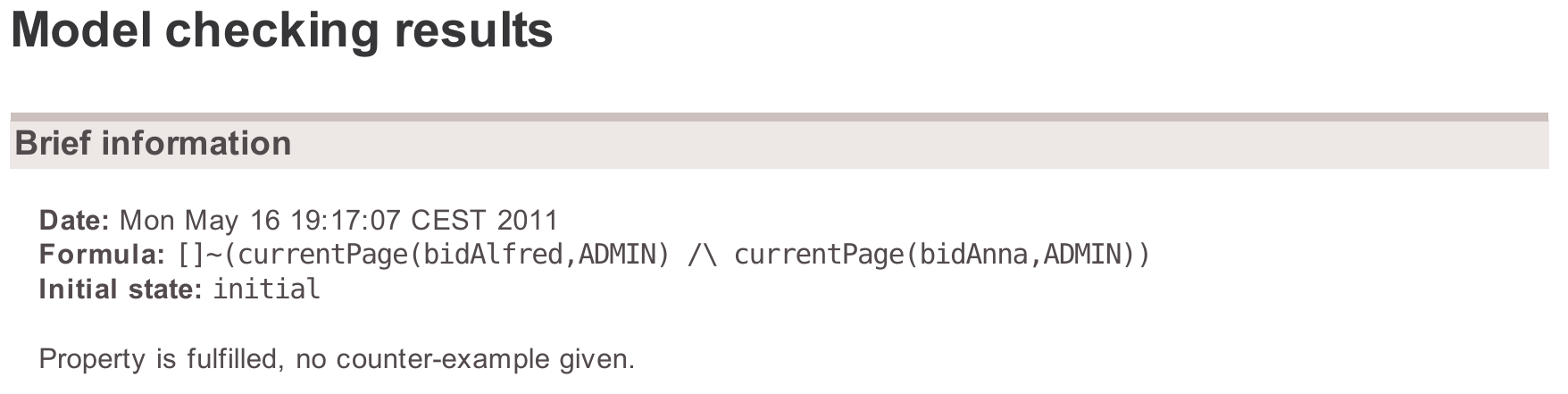}
}
\caption{Snapshot of the {\sc Web-TLR} System for the case of no counter-examples.}\label{fig:no-counterexample}
\end{figure}

Finally, by using {\sc Web-TLR} again we get the outcome {\sf ``Property is fulfilled, no counter-example given''}, which guarantees that now the Web application satisfies Property \ref{property}.  
Figure~\ref{fig:no-counterexample} shows a snapshot of {\sc WEB-TLR} for the case when a property is fulfilled.

\section{Conclusions}\label{sec:conclu}

We have presented an extension of {\sc Web-TLR} which provides a backward trace-slicing facility that eases the interactive debugging of Web applications. 
The proposed slicing technique greatly reduces the size of the counterexample traces thus making their analysis feasible even in the case of complex, real-size Web systems. 
The proposed slicing technique allows that
the size of the counterexample trace  be greatly reduced, 
which makes their analysis feasible even in the case of complex, real-size Web systems.

The tool provides a complete description of all events and locations that are involved in a particular error scenario, and 
the user can analyze different error scenarios in an incremental, step-by-step manner.

We have tested our tool on several complex case studies that are available 
at the \textsc{Web-TLR} Web page and within the distribution package, including a Webmail application, a producer/consumer system and a sophisticated
formal specification of a fault-tolerant communication protocol.
The  results obtained are very encouraging and show  impressive reduction rates in all cases,  ranging from $90\%$ to $95\%$. 
Moreover, sometimes the trace slices are so small that they can be easily inspected by the user who can keep a quick eye on what's going on behind the scenes.   A thorough experimental evaluation of our trace slicing technique can be found in \cite{ABER11tr}.

Currently, Web-TLR only supports the formal modeling and verification of traditional Web applications. 
As future work, we plan to extend our framework to deal with more sophisticated Web systems based on Web service architectures (e.g., those conforming to the REST framework \cite{FT02}). 



\end{document}